\documentclass[preprint, amsmath,amssymb,prb,superscriptaddress,floatfix]{revtex4-2}
\usepackage[utf8]{inputenc}
\usepackage{color}
\usepackage{graphicx}
\usepackage[caption=false]{subfig}
\usepackage{placeins}
\usepackage[export]{adjustbox}
\usepackage[version=4]{mhchem}
\usepackage{multirow}
\usepackage{tabularx}
\graphicspath{{Figures/}}
\usepackage{xspace}
\newcommand{\iang}{~\AA$^{-1}$\xspace}
\usepackage{hyperref}

\begin{document}
\title{Diffuse spin waves and classical spin liquid behavior in the kagom\'e antiferromagnet chromium jarosite, \texorpdfstring{\ce{KCr3(OD)6(SO4)2}}{KCr3(OD)6(SO4)2}}

\author{Sofie Holm-Janas}
\affiliation{Nanoscience Center, Niels Bohr Institute, University of Copenhagen, 2100 Copenhagen, Denmark}\affiliation{Laboratory for Quantum Magnetism, Institute of Physics, EPFL, 1015 Lausanne, Switzerland}
\affiliation{Department of Physics, Technical University of Denmark, Fysikvej, 2800 Kongens Lyngby, Denmark}

\author{Sidse L.~Lolk}%
\affiliation{Nanoscience Center, Niels Bohr Institute, University of Copenhagen, 2100 Copenhagen, Denmark}

\author{Anders~B.~A.~Andersen}%
\affiliation{Department of Physics, Chemistry and Pharmacy, University of Southern Denmark, 5230 Odense M, Denmark}

\author{Tatiana~Guidi}
\affiliation{ISIS Facility, Rutherford Appleton Laboratory, OX11 0QX Didcot, United Kingdom}

\author{David~Voneshen}
\affiliation{ISIS Facility, Rutherford Appleton Laboratory, OX11 0QX Didcot, United Kingdom}
\affiliation{Department of Physics, Royal Holloway University of London, TW20 0EX, United Kingdom}

\author{Ivica~{\v Z}ivkovi{\' c}}
\affiliation{Laboratory for Quantum Magnetism, Institute of Physics, EPFL, 1015 Lausanne, Switzerland}

\author{Ulla~Gro~Nielsen}%
\affiliation{Department of Physics, Chemistry and Pharmacy, University of Southern Denmark, 5230 Odense M, Denmark}

\author{Kim~Lefmann}%
\affiliation{Nanoscience Center, Niels Bohr Institute, University of Copenhagen, 2100 Copenhagen, Denmark}

\begin{abstract}
    The dynamics of the $S=3/2$ kagom\'e antiferromagnet chromium jarosite, \ce{KCr3(OD)6(SO4)2}, was studied using high-resolution neutron time-of-flight spectroscopy on a polycrystalline sample with a nearly stoichiometric magnetic lattice (2.8(2)$\%$ Cr vacancies). Neutron spectroscopy reveals diffuse spin wave excitations in the ordered phase with an incomplete gap and significant finite life-time broadening as well as a pronounced kagom\'e zero mode. Using linear spin wave theory, we estimate the exchange couplings. The system is highly two-dimensional with the leading nearest-neighbor coupling being $J_1 = 0.881$~meV. Above $T_N$ diffuse excitations from the classical spin liquid regime dominate. We model the $Q$- and energy-response separately, and show that in both the ordered phase and the classical spin liquid regime they are strongly coupled.
\end{abstract}

\maketitle

\section{Introduction}
Emergent phenomena occur within geometrically frustrated spin systems, where competing interactions suppress the normal magnetic ordering of the spins, and create a macroscopic ground state degeneracy with a flat energy landscape for excitations. In certain systems, this allows exotic order and disorder phenomena to emerge, 
such as the spin liquid \cite{Balents2010,Wen2019,Savary2017,Knolle2019}. In these states the spins fluctuate in a correlated liquid-like manner. In quantum spin liquids, the system eludes order down to temperatures of absolute zero. Furthermore, they are predicted theoretically to show long-range entanglement at 0~K \cite{Savary2017}. 
However, the experimental verification of a quantum spin liquid have proven contentious, also due to inherent problems with structural defects in synthesized samples \cite{Chamorro2021}. In contrast, the classical spin liquids are more prolific but studied comparatively less. Here, the frustration only partially suppresses the magnetic ordering, so that the system still orders at finite temperatures. This prevents long-range entanglement at any temperature. The interplay between order and disorder in the classical spin liquid regime, sometimes also termed the cooperative paramagnetic phase, gives rise to exotic liquid-like fluctuations at intermediate temperatures due to the frustrated interactions \cite{Knolle2019}.\\

\begin{figure}[htb]
    \includegraphics{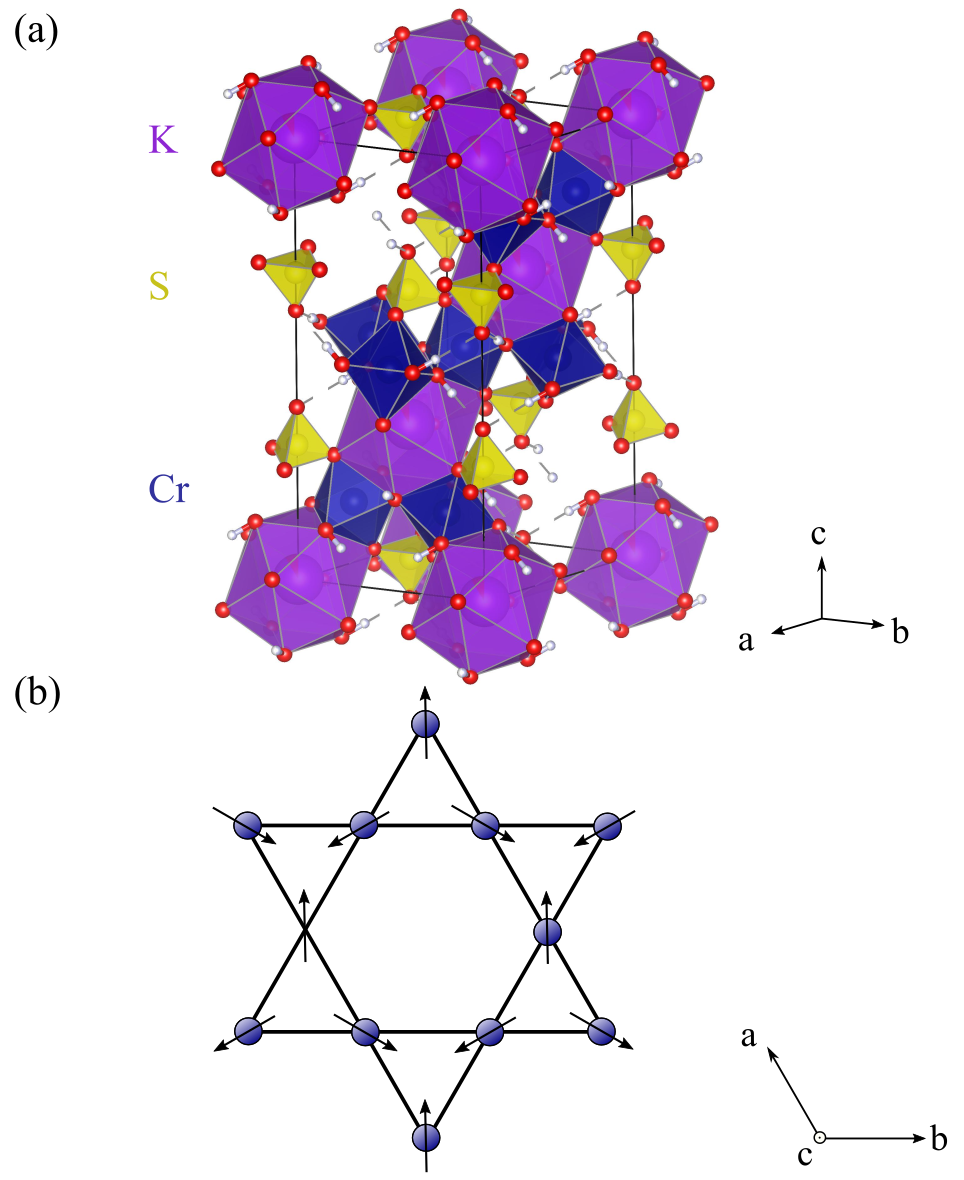}
    \caption{(a) Crystal structure of Cr-jarosite. (b) Magnetic ground state in the 120$^\circ$ configuration in the two-dimensional chromium kagom\'e planes.}
    \label{fig:crystal}
\end{figure}

The kagom\'e antiferromagnet has taken a center stage in the study of geometrically frustrated spin liquids. This is due to the two-dimensional nature and the low connectivity of $z=4$ nearest neighbors in the kagom\'e geometry as illustrated in Fig.~\ref{fig:crystal}(b), which combined with the frustration from the antiferromagnetic couplings creates a macroscopic ground state degeneracy. The presence of dispersionless zero-energy modes in the ground state manifold prevents long-range order even at $T=0$, both in the classical and quantum spin system \cite{Mendels2011}. These zero-energy modes (dubbed {\em kagom\'e zero modes}) are a special property of the kagom\'e geometry, and originate from the fact that coherent local rotations of the spins in two sublattices around the third may occur at no energy cost \cite{Mendels2011}. In most physical realizations of the kagom\'e antiferromagnet the presence of additional, non-dominant interactions will cause the systems to order magnetically at finite temperatures. These additional interactions will cause the zero-energy mode to be lifted up to finite energy transfers, which has the fortunate consequence that they can be observed more easily in neutron experiments. In spite of this, the kagom\'e zero mode has only been observed so far in three classical kagom\'e antiferromagnets \cite{Matan2006,Hayashida2020,Kermarrec2021}.

One of the most studied classical kagomé antiferromagnetic materials is the jarosite family of magnets with the general chemical composition \ce{AM3(OH)6(SO4)2}, where A is a monovalent ion (\ce{Na+,K+,H3O+,Ag+,NH4+}) and \ce{M} is a trivalent ion, which may be magnetic (\ce{Fe^3+, Cr^3+, V^3+}). Jarosites crystallize in the $R\bar{3}m$ space group with magnetic ions positioned in stacked kagom\'e planes, as shown in Fig.~\ref{fig:crystal}. Jarosite originally refers to the $S=5/2$ iron mineral \ce{KFe3(SO4)2(OH)6} (henceforth Fe-jarosite), which is a frustrated antiferromagnet and  orders at $T_N = 65$~K. Fe-jarosite is well-described as a Heisenberg antiferromagnet with a Dzyaloshinskii-Moriya (DM) interaction, which causes a spin canting within the kagom\'e layers \cite{Grohol2003}. The magnetic properties of Fe-jarosite have been studied extensively \cite{Grohol2003,Nocera2004,Grohol2005,Matan2006,Matan2011,Fujita2012,Klein2018,Mook2019}, both in powders and single crystal specimens. Exchanging the magnetic ion to \ce{V^3+} lifts the frustration and causes ferromagnetic ordering in V-jarosite \cite{Grohol2002,Papoutsakis2002}. However, markedly fewer studies have addressed the magnetic behavior of the $S=3/2$ Cr-analogue, \ce{KCr3(OH)6(SO4)2} (Cr-jarosite). Cr-jarosite is also a frustrated antiferromagnet \cite{Janas2020,Okuta2011,Okubo2017}, and is the focus of this paper. 

Studies of jarosites, and in particular of Cr-jarosite, have suffered significantly from the difficulty in synthesizing pure and stoichiometric compounds. This is a general and significant problem for many types of frustrated materials, in which structural disorder and defects complicate the analysis of the weak magnetic behavior \cite{Chamorro2021}. However, this was improved for jarosites by introduction of the redox-hydrothermal synthesis in the early 2000's \cite{Grohol2002, Grohol2003,Nocera2004} for M = Fe and V.  For jarosites, the two main defects are a vacancy on the magnetic B site, and mixed occupancy on the A site, which may or may not be correlated\cite{Nielsen2008, Janas2020}. An M site vacancy will locally destroy the kagom\'e lattice by lifting the magnetic frustration, leading to local antiferromagnetic interactions and an A site vacancy \cite{Nielsen2011}. It is an experimental fact that while Fe-jarosites with A = \ce{K+, NH4+,Na+,Ag+} will order magnetically, jarosites with A = \ce{H3O+} instead show spin glass behavior. Thus, the A site cations also affect the magnetic properties, but currently an understanding of the mechanism is lacking \cite{Wills2001xyz,Grohol2007,Bisson2008,Wills2011,Spratt2014}. Although the redox-hydrothermal synthesis has improved the stoichiometry, our recent study of Cr-jarosite prepared using this approach revealed significant (2-27\%) substitution of K by \ce{D3O+} and \ce{D2O} species \cite{Janas2020}.

While the static magnetic behavior of Cr-jarosite has previously been contested (see e.g. \cite{Townsend1986,Keren1996,WillsPhD,Wills2001}), all recent results \cite{Janas2020,Okubo2017,Okuta2011} point to the same overall conclusions: Cr-jarosite orders antiferromagnetically at $T_N \approx 4$~K and has a Curie-Weiss temperature of $\theta_\textrm{CW} \approx -62 K$, meaning it has a frustration ratio of $f=|\theta_\textrm{CW}|/T_N\approx 16$ \cite{Janas2020}. Below $T_N$ it assumes a 120$^\circ$ noncollinear ground state with a small ferromagnetic canting in each layer due to a small DM interaction, see Fig.~\ref{fig:crystal}(b) \cite{Janas2020,Okubo2017,Okuta2011}. The kagom\'e layers are coupled ferromagnetically, causing the spins to cant the same way in each layer which adds up to a small positive magnetization. 
However, the excitations in Cr-jarosite have only previously been investigated in a single study made before the advent of the redox-hydrothermal synthesis on a sample with 26\% Cr vacancies and an undisclosed amount of K vacancies \cite{Lee1997}. Such a high Cr vacancy content significantly affects the magnetism of the sample, as 33 \% Cr vacancies completely lift the magnetic frustration, making the study not representative of stoichiometric Cr-jarosite.

In this paper we will study the classical spin liquid compound Cr-jarosite, and remedy the lack of excitation studies by use of neutron spectroscopy on a well-defined sample with a stoichiometric magnetic lattice. The vacancy levels were quantified using \ce{^{2}H} MAS NMR, which is highly sensitive to local environments within the crystal lattice, and therefore ideal as a tool for defect quantification. In the ordered state below $T_N$, the excitation spectrum can be described by dampened spin waves which are modeled using linear spin-wave theory to extract estimates of the exchange interactions. In the classical spin liquid regime between above $T_N$ the excitations are of a short-ranged nature. By modeling both the energy- and $Q$-dependence of the excitations we link the behavior to that of a classical spin liquid.

\section{Experimental details}
Polycrystalline, deuterated Cr-jarosite, \ce{KCr3(OD)6(SO4)2}, was prepared by the redox-hydrothermal method \cite{Grohol2003} with combined mass of 6.9 g. The purity of the sample as well as each batch prior to mixing was assessed by powder X-ray diffraction (PXRD) and \ce{^{2}H} MAS NMR. The synthesis and characterization followed our recently reported  procedure \cite{Janas2020} with further details are given in appendix \ref{app:synth}.
\\

\ce{^{2}H} MAS NMR single-pulse and rotor-synchronized Hahn-echo (90$^{\circ}$–$\tau$–180$^{\circ}$–$\tau$–acquisition) spectra were performed on an Agilent INOVA spectrometer at 14.1 T equipped with a 1.6 mm triple-resonance MAS probe using several spinning speeds in the range 30-35 kHz for unambiguous identification of the isotropic regions, see \cite{Janas2020}. The echo delay ($\tau$) was one rotor period. The \ce{^{2}H} MAS NMR spectra were referenced relative to TMS using \ce{D2O} ($\delta_{\rm iso} $ = 4.6 ppm) as secondary reference. The \ce{^{2}H} MAS NMR data were processed in the MestReNova software (ver.\ 12.0.1). Spectral deconvolution and fitting of the chemical shift anisotropy (CSA) parameters ($\Delta$ and $\eta_{\sigma}$) and quadrupole coupling ($C_Q$ and $\eta_Q$), were performed in the ssNake software \cite{ssnake} (ver.\ 1.2), see \cite{Janas2020}. The CSA parameters are reported in the Haeberlen convention \cite{Haeberlen}, where the Euler angle $\beta$ relates the principal axis of the CSA and quadrupole tensor \cite{skibsted1992}.\\

Susceptibility data was measured between 5 K and 370 K using a Quantum Design MPMS-5 SQUID. Low-temperature measurements of the magnetization was made using a Quantum Dynacool Physical Property Measurement System (PPMS).\\

Neutron spectroscopy was performed at the ISIS neutron facility (UK). Initially, the excitations were studied using the time-of-flight spectrometer MARI using a He-4 cryostat \cite{Taylor1990}, which revealed a need for better low-energy resolution. Thus, we proceeded to the time-of-flight spectrometer LET (ISIS) \cite{Bewley2011}, where the powder was loaded into a hollow cylindrical aluminum canister in a helium atmosphere, and placed in an orange-type cryostat. Repetition rate multiplication energies of 2.20 meV, 3.70 meV, 7.60 meV, and 22.20 meV provided simultaneous access to different parts of reciprocal space. 
For investigations of Bragg reflections, $E_i = 2.20$~meV was used for slightly better Q-resolution, while $E_i = 3.70$~meV data is used for all other investigations as the best compromise between resolution and $(Q,\omega)$ coverage of the excitations. Data was taken at 9 different temperatures between 1.8 K and 100 K. The initial data reduction was performed with Mantid \cite{Arnold2014}, and data analysis was performed using Horace \cite{Ewings2016}. Data modeling of spin waves was performed using a linear spin-wave theory based on the Holstein-Primakoff framework as implemented in the SpinW package \cite{Toth2015}. 

\section{Characterization}
PXRD and SEM-EDS was used to determine the purity and chemical composition of the combined sample. The PXRD diffractograms (see appendix \ref{app:PXRD}) showed only reflections from chromium jarosite, and are in excellent agreement with our recent study \cite{Janas2020}. Thus, no crystalline impurities were identified by PXRD. It is noted that the (107) reflection shows a complex line shape indicative of partial hydronium substitution on the K site, as discussed in detail in Ref.~\cite{Janas2020}.\\

The concentration of defects were quantified by \ce{^{2}H} MAS NMR spectroscopy for the combined sample following the same procedure as earlier reported \cite{Janas2020}. The Cr and K defects have previously been shown to be correlated, and can be summarized in the general chemical formula as \cite{Janas2020},
\begin{align}
\ce{[K_{1-x-y}(D_nO)_{y+x}] Cr_{3-x}(SO4)2(OD)_{6-4x}(OD2)_{4x}},
\end{align}
where \ce{D_nO} refers to the total concentration of deuterium species on the A site, which is the sum of \ce{D2O} and \ce{D3O+.} This highlights that \ce{Cr} vacancies also cause \ce{K} vacancies, but \ce{K} vacancies can also occur on their own.
The concentration of the defects may be determined by deconvolution of the \ce{^{2}H} MAS NMR spectrum (see appendix \ref{app:NMR}). The Cr-vacancies are then determined from the integrated intensities of the appropriate \ce{^{2}H} NMR resonances \cite{Nielsen2008}:
\begin{equation}
4x = 6 \frac{\frac{1}{2} I(\ce{CrOD2})} {I(\ce{Cr2OD})+\frac{1}{2}I(\ce{CrOD2})}. \label{eq:x}
\end{equation}
The average number, $n$, of deuterons on the A-site may similarly be estimated by \cite{Nielsen2008}
\begin{equation}
n = 6 \frac{I(\ce{D_nO})}{I(\ce{Cr2-OD}) + \frac{1}{2}I(\ce{Cr-OD2})}, \label{eq:n}
\end{equation}
where values of $n=3$, 2, or 0 corresponds to the A-site being fully occupied by \ce{D3O+}, \ce{D2O}, or \ce{K+} respectively. 

\begin{figure}[htb]
    \includegraphics[width=0.4\textwidth]{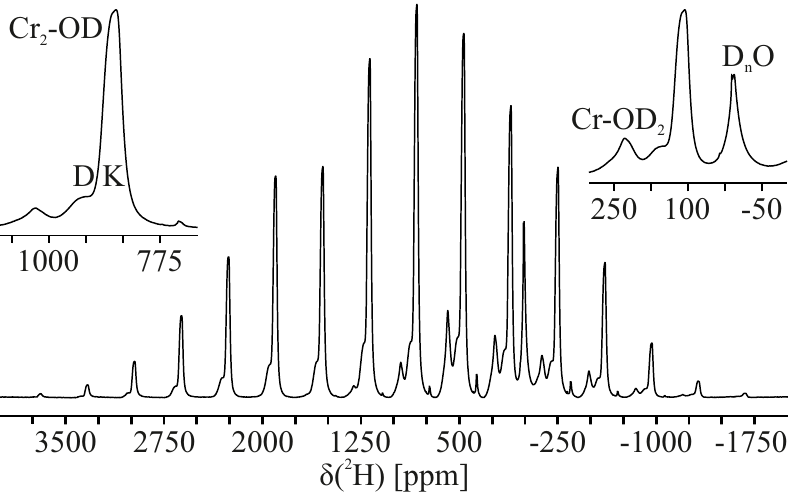}
    \caption{\ce{^{2}H} MAS NMR spectrum (Hahn Echo) of polycrystalline Cr-jarosite recorded at 14.1 T with 33 kHz spinning speed. The left insert illustrates the isotropic region for the Cr\ce{_{2}}-OD (D) and Cr\ce{_{2}}-OD (K) resonances. The right insert shows the isotropic region for the Cr-OD\ce{_{2}} and D\ce{_{n}}O resonances.} 
    \label{fig:NMRspectrum}
\end{figure}

The \ce{^{2}H} MAS NMR spectrum (Fig.~\ref{fig:NMRspectrum}) contains the spinning side band manifold for four resonances located at $\delta_{iso} $ = 872(10), 828(5), 229(5), and 7(2) ppm, which are assigned to \ce{Cr2-OD} (D), \ce{Cr2-OD} (K), \ce{Cr-OD2} and \ce{D_{n}O}, respectively, using earlier reported data \cite{Grube_2018, Janas2020}. The \ce{Cr2-OD} resonances stem from stoichiometric Cr-jarosite with either \ce{D_{n}O} or \ce{K+} on the A-site. The \ce{Cr-OD2} resonance stem a Cr next to a Cr vacancy (defect), whereas the \ce{D_{n}O} resonances contain the overlapping resonances from \ce{D_{n}O} on the A-site and surface adsorbed \ce{D2O}. Moreover, a negligible amount ($<1$\%) of a sharp \ce{D_{n}O} resonance at 11(1) ppm from an unknown impurity were observed. Thus, the \ce{^{2}H} MAS NMR spectra also confirm partial substitution of hydronium in the sample. 

Based on these values, eq.~\ref{eq:x} yields $x = 0.085(1)$, corresponding to 2.8(2)\% Cr vacancies (97.2(2) \% occupancy on the magnetic lattice). Eq.~\ref{eq:n} yields $n = 0.41(9)$ and thereby a K occupancy of 80-86 \%, where the upper and lower boundary corresponds to substitution by only \ce{D3O+} or \ce{D2O} on the A-site, respectively. Thus, the analysis of the combined sample by PXRD, SEM/EDS, and \ce{^{2}H} MAS NMR confirms the formation of Cr-jarosite. Based on the quantified level of defects, the magnetic kagom\'e lattice of Cr ions is almost perfectly intact. The K-defects are at a quantified level and match our earlier study\cite{Janas2020}.\\

\begin{figure}[tb]
    \includegraphics{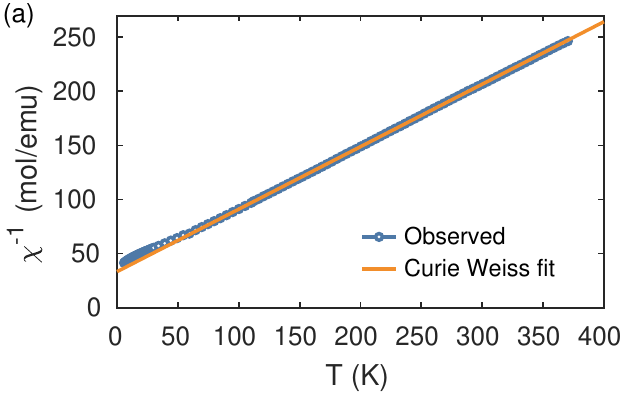}
    \includegraphics{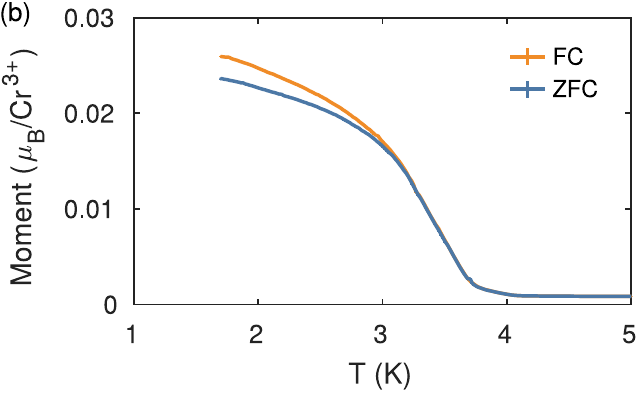}
    \caption{Magnetization of Cr-jarosite. (a) Inverse molar susceptibility measured in 3~T fitted with the Curie-Weiss law. (b) Magnetization in 0.02~T field.} 
    \label{fig:MPMS}
\end{figure}

The susceptibility was measured from 5 K -- 370 K in applied fields of 3~T, see Fig.~\ref{fig:MPMS}(a). Fitting the data with the Curie-Weiss law and assuming $g=1.97$ as determined from electron spin resonance (ESR) \cite{Okubo2017}, yields a Curie-Weiss temperature of $\theta_{\rm CW} \approx -58$~K, in close agreement to previously obtained values \cite{Janas2020,Okuta2011}.
Assuming that the nearest-neighbor interactions within the kagom\'e plane dominates, a mean-field estimate for the in-plane exchange coupling can be obtained as
\begin{align}
J_\mathrm{MF} = \frac{3 k_B |\theta_{\rm CW}|}{z S(S+1)} \approx 1.0~\mathrm{meV},
\end{align}
where $z=4$ is the number of nearest-neighbors. 

The field cooled (FC) and zero-field-cooled (ZFC) magnetization of the sample for low temperatures in 0.02~T is shown in Fig.~\ref{fig:MPMS}(b), which shows an antiferromagnetic phase transition with weak ferromagnetic canting in the region $3-4$~K. The FC and ZFC magnetization show a small difference at low temperatures, consistent with the hysteresis in the system due to the canted antiferromagnetic moment \cite{Janas2020}. At 1.7~K, the magnetization is measured as 0.024~$\mu_B$/\ce{Cr^3+} in the powdered sample, corresponding to 0.05~$\mu_B$/\ce{Cr^3+} in a single crystal with magnetic field along the $c$ axis, matching previous results \cite{Okuta2011,Janas2020}. Assuming a saturated magnetization of 3.87$~\mu_B$ for \ce{Cr^3+} in an octahedral ligand field, this corresponds to a spin canting angle of approximately $\alpha = \sin^{-1}(M/M_s)\approx 0.7^\circ$. \\

\begin{figure}[tb]
    \includegraphics{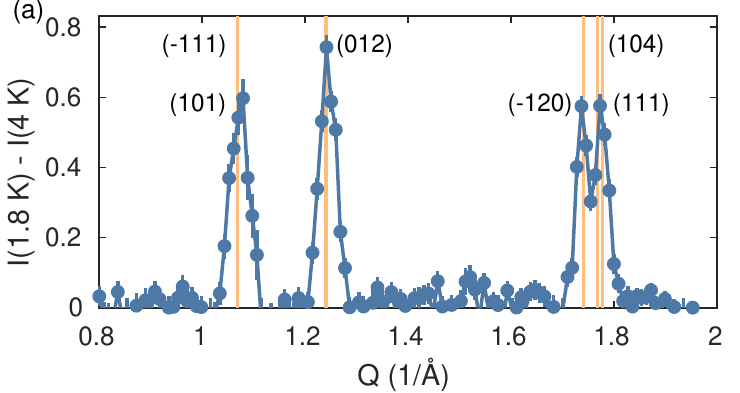}
    \includegraphics{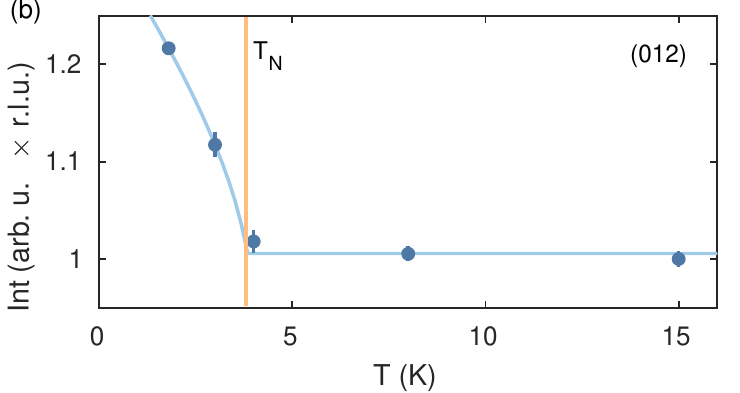}
    \caption{Elastic scattering from Cr-jarosite measured on LET. Data is obtained using $E_i=2.20$~meV, and integrated in the energy transfer range $\Delta E =\pm 0.08$~meV. (a) Difference between the elastic line at 4~K and 1.5~K with indexed magnetic Bragg peaks. (b) Temperature versus integrated intensity for the purely magnetic (012) reflection around $Q=1.25$~\iang. The blue line is a guide-to-the-eye power law fit with $T_N = 3.8$~K.} 
    \label{fig:Bragg}
\end{figure}

\begin{figure}[htb]
    \includegraphics{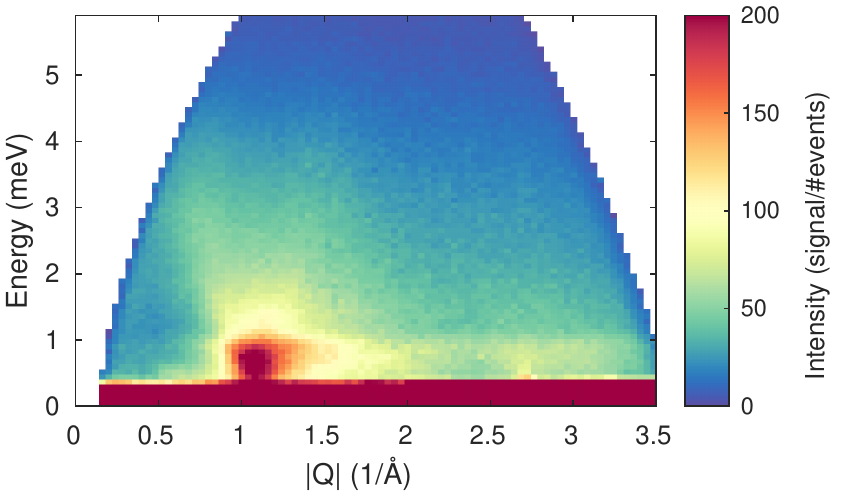}
    \caption{Excitations at 1.8~K in Cr-jarosite measured using neutron spectroscopy at LET with $E_i=7.60$~meV.} 
    \label{fig:Ei7-60}
\end{figure}

\begin{figure*}[thb]
    \captionsetup[subfloat]{labelformat=empty}
    \subfloat[][]{\includegraphics[valign=t]{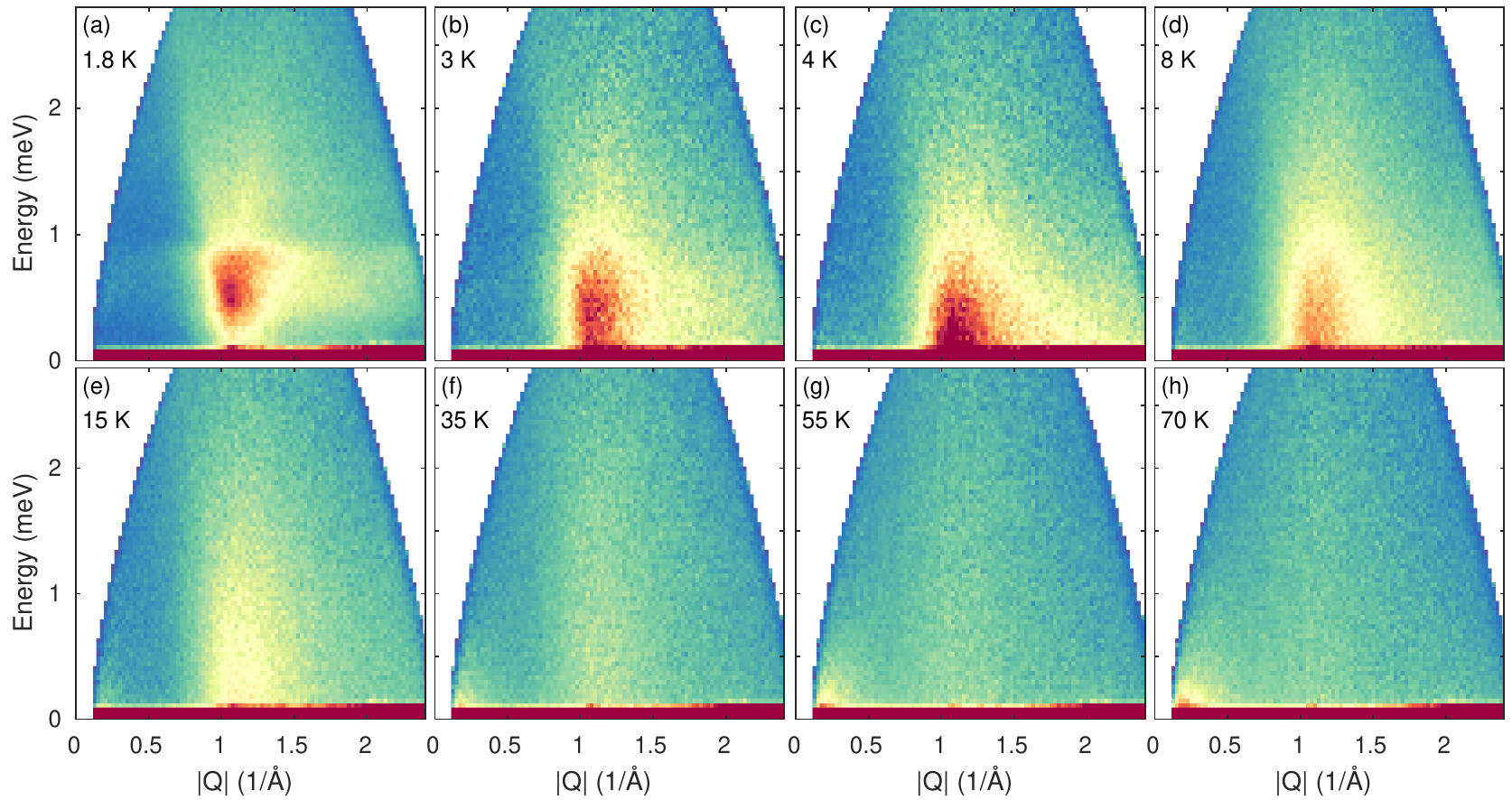}}
    \subfloat[][]{\includegraphics[valign=t,trim = {4.26cm 0 0 0},clip]{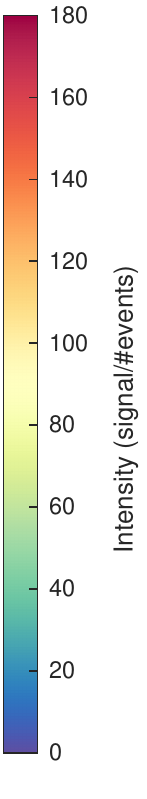}}
    \caption{Excitation spectra in Cr-jarosite measured by inelatic neutron scattering at LET for various temperatures using $E_i=3.70$~meV.}
    \label{fig:allT}
\end{figure*}
The magnetic and structural Bragg reflections in Cr-jarosite were investigated to ascertain the magnetic ordering (Fig.~\ref{fig:Bragg}). This elastic scattering data was obtained from the LET data with $E_i = 2.20$~meV by taking constant-energy cuts around the elastic line. Figure \ref{fig:Bragg}(a) shows the intensity difference between 1.8~K and 4~K, and highlights the purely magnetic Bragg reflections. All reflections can be indexed by hand using a magnetic unit cell identical to the structural. This is consistent with a ground state with the 120$^\circ$, $\mathbf{q}=0$ spin arrangement with a small ferromagnetic canting out of the kagom\'e plane. Note that no actual Rietveld refinement of the structure is possible due to the poor resolution and $Q$ coverage of spectrometer data compared to actual diffractometer data. Fig.~\ref{fig:Bragg}(b) shows the low-temperature evolution of the intensity of the magnetic (012) reflection. This is consistent with both the susceptibility measurements in Fig.~\ref{fig:MPMS} and the results of $T_N=3.8$~K from our previous neutron investigation\cite{Janas2020} for samples with a similar vacancy profile.

\section{Magnetic excitations}
The excitation spectra for the lowest measured temperature can be seen in Fig.~\ref{fig:Ei7-60} for data using $E_i=7.60$~meV, showing that all relevant magnetic excitations appear at energies below $4$~meV. Thus the analysis of the excitations will focus on the lower energies accessed with $E_i=3.70$~meV.

The temperature development of the excitation spectra can be seen in Fig.~\ref{fig:allT}.
In the magnetically ordered phase at $T=1.8$~K $\approx T_N/2$ (upper left corner), distinct spin wave excitations can be seen. The data reveal a continuum of high intensity scattering centered at $Q \approx 1.1$\iang and $\hbar \omega = 0.7$~meV with indications of a gap. The spin wave branches continue up to roughly 4~meV as seen in Fig.~\ref{fig:Ei7-60}. Additionally, a gapped, broad, and approximately flat mode exists at $\hbar \omega = 0.4 - 0.9$~meV extending up to $Q \approx 3.5$\iang. As shall be shown later, this flat mode is a realization of a kagom\'e zero mode lifted to finite energies due to the DM interaction.

\begin{figure}[tb]
    \includegraphics{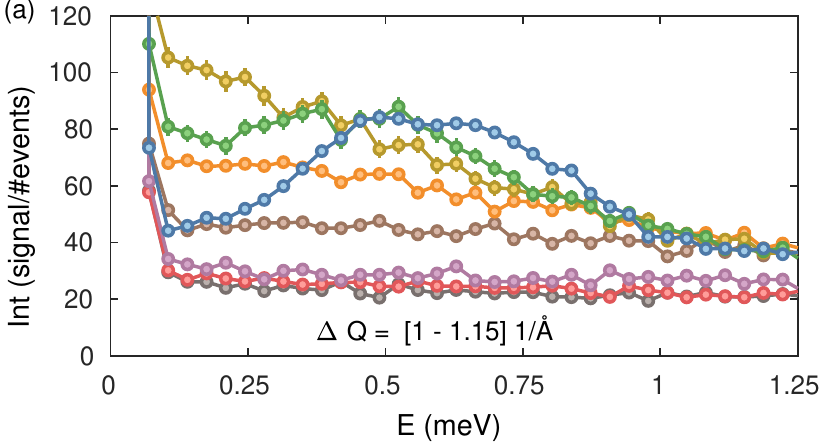}
    ~
    \includegraphics{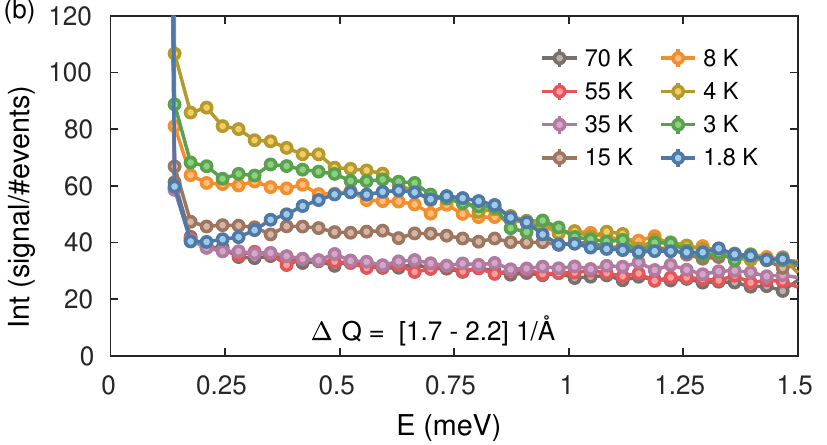}
    \caption{Constant-Q cuts through the excitations integrated over energy for various temperatures. (a) Cut through the intense excitation at $Q \approx 1.07$~\iang with $E_i = 2.20$~meV. (b) Cut through the flat mode at $Q \approx 2$~\iang with $E_i = 3.70$~meV.}
    \label{fig:cuts}
\end{figure}

Beginning from the low-temperature data, the intensity moves towards lower energies as the temperature increases from 1.8~K through $T_N$, and the spectra starts to lose the distinct spin waves. Above $T_N$, the excitations remain as a distinct feature with intensity above $Q \approx 1.1$\iang and no gap to the elastic line visible in the $E_i=3.70$~meV data. This feature remains visible up to $\sim15-35$~K. Some broader, increased intensity is also present at $55 - 70$~K, which can be recognized as para\-magnetic scattering that is practically constant within the energy scale of the system and decreases in a way consistent with the magnetic form factor for larger $Q$.
From 35~K to 70~K, increased low-energy intensity for very low $Q$ appears. Due to its temperature-dependence, we believe this to be related to structural small-angle scattering, and have not included this further in the analysis.

The distinct excitations in the range $\sim 4-15$~K are magnetic in nature, as they are continuously connected to the spin wave excitations at low temperatures, disappear at higher temperatures, and originate from the position of the strong magnetic Bragg reflections. Given their diffuse nature, they indicate short-ranged magnetic correlations within the classical spin liquid regime. \\

The existence of a gap in the excitation spectrum is illustrated in Fig.~\ref{fig:cuts}, where (a) shows a cut through the strong continuum centered at $Q\approx 1.1$\iang, while (b) shows a cut through the lifted kagom\'e zero mode at higher $Q$. For the spin wave continuum above the (101) reflection, the gap appears incomplete even at 1.8~K, although it may open fully at lower temperatures. The gap is still vaguely present in the 3~K data, while being completely closed at 4~K. The gap from the elastic line to the kagom\'e zero mode around $Q\approx 2$\iang (b) is, however, a complete gap with a value of $\sim 0.4$~meV at 1.8~K. As the temperature increases, this gap also closes.

\subsection{Spin waves}
\label{sec:spinwaves}
\begin{figure}[tp]
    \includegraphics{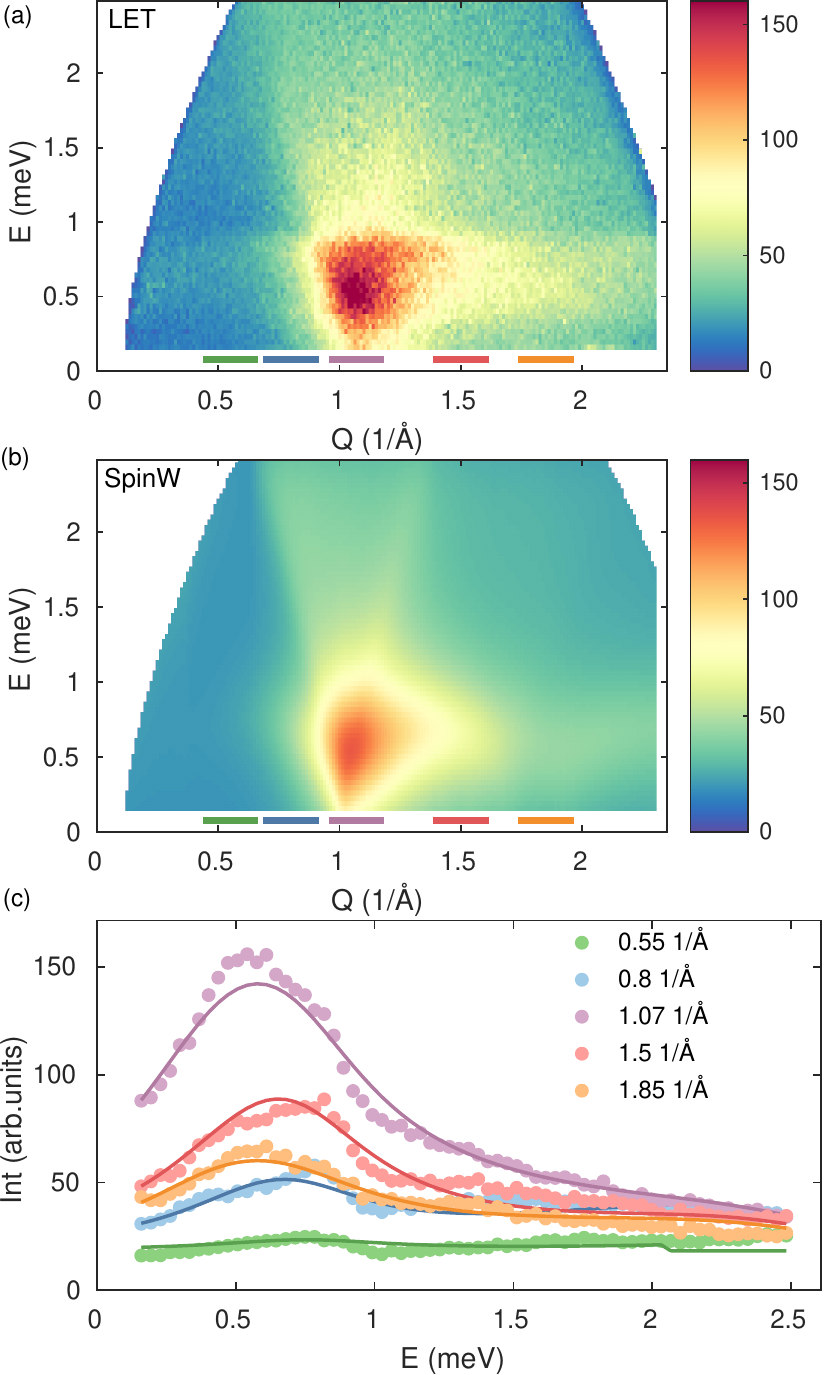}
    \caption{Spin wave excitations in Cr-jarosite. (a) Spin waves measured with LET at $T=1.5$~K. (b) Resulting  SpinW spectra of best fit. (c) Simultaneous SpinW fit of $Q$-cuts through data as described in the text. The positions and widths of the Q-cuts are indicated in subfigs (a-b) with horizontal colored lines.} 
    \label{fig:SW}
\end{figure}

Figure \ref{fig:SW}(a) shows the measured spin wave sprectrum well within the ordered phase at $T = 1.8$~K. We will analyze the excitations in the ordered state with linear spin wave theory to obtain estimates for the exchange couplings. The spin wave excitations in Fig.~\ref{fig:SW}(a) lack any clear van Hove singularities, and appear to be significantly broadened. This may either stem from finite lifetimes of the spin waves, which would cause a broadening in energy, or from co-existing, persistent short-ranged correlations, which would cause a broadening in \textit{Q}. Both of these effects are inherent to frustrated lattices, but may also be exacerbated by the K disorder in the sample, which may also cause diffuse scattering. No matter the precise origin, the effect on the data will be similar. For technical reasons, related to calculating powder averages using SpinW, we will assume an energy broadening of the spin waves as stemming from finite lifetimes.  \\

We assume the magnetic ground state to be the 120$^\circ$, $\mathbf{q}=0$ spin structure with equal canting in each layer.  We assume the magnetic system to be described by the Hamiltonian,
\begin{align}
\begin{split}
\mathcal{H}=&\sum_{\mathrm{nn}}\left[J_{1} \vec{S}_{i} \cdot \vec{S}_{j}+\vec{D}_{i j} \cdot \vec{S}_{i} \times \vec{S}_{j}\right]\\
+&\sum_{\mathrm{nnn}} J_{2} \vec{S}_{k} \cdot \vec{S}_{l} +\sum_{\mathrm{n_\perp n_\perp}} J_{\perp} \vec{S}_{m} \cdot \vec{S}_{n} ,
\label{eq:jaroham}
\end{split}
\end{align}
where $J_1$ and $J_2$ are the in-plane nearest- and next-nearest neighbor antiferromagnetic exchange couplings, while $J_\perp$ is the interlayer nearest-neighbor coupling. $J_\perp$ is constrained to be ferromagnetic due to the ordered state having the same in-plane spin arrangement in neighboring layers. $\vec{D}_{i j} = \left[\sqrt{3} D_{y}(i, j), D_{y}(i, j), D_{z}(i, j)\right]$ is the symmetry-allowed DM interaction, where we also assume both $D_y,D_z < 0$. While the exact role of the exchange parameters was found analytically by \citet{Yildirim2006}, their overall contribution can be summarized as: $J_1$ controls the bandwidth of the high-intensity, higher-energy spin wave, $D_y$ and $D_z$ control the center of the flat mode and causes a canting of the spins, and $J_2$ causes dispersions of the flat mode, such that $D_y$, $D_z$, and $J_2$ together control the bandwidth of this mode. $J_\perp$ is not included in the analytical treatment, but small $J_\perp$ also induces dispersions in the flat mode.\\
The spin wave dispersions were calculated using the SpinW package, and then powder-averaged to compare with experimental data. 
In order to account for the experimental resolution along $Q$, the powder-averaged spectra were convoluted with a Gaussian with full-width half-maximum (FWHM) of $\Delta Q = 0.1$~\iang as estimated from the width of the Bragg reflections. In order to account for the overall smearing modeled by the finite lifetimes of the excitations, an additional variable parameter is introduced by convolution of the spectra with a Lorentzian with FWHM $\Delta E$. 
We used a constrained particle swarm algorithm to perform $\chi^2$ fitting of constant-$Q$ cuts through data. To capture most features of the spin waves while also keeping run times manageable, we simultaneously fitted 5 cuts through $Q$ at $Q = 0.55, 0.8, 1.07, 1.5, 1.85$~\iang with width $dQ = 0.1$~\iang, as illustrated in Fig.~\ref{fig:SW}. 
We fitted the data using 7 free parameters: $J_1$, $J_2$, $J_\perp$, $D_y$, $\Delta E$, as well as a common intensity scale and constant background. We assumed that the DM interaction had the same $D_z/D_y = 0.255$ ratio as the values obtained by \citet{Okubo2017} based on antiferromagnetic resonance measurements, but we allowed the strength to vary. The next-nearest neighbor $J_2$ was included in the fit since it significantly improved the reduced $\chi^2$ of the fit. The spin canting was optimized for each choice of the exchange parameters to minimize the ground state energy.\\

The resulting best fit can be seen in Fig.~\ref{fig:SW}. Fig.~\ref{fig:SW}(a) shows the experimental spectrum, Fig.~\ref{fig:SW}(b) shows the resulting SpinW spectrum, and Fig.~\ref{fig:SW}(c) shows the five 1D cuts that the fit was made to. The best fit has the parameters $J_1 = 0.881$~meV, $J_2=0.015$~meV, $J_\perp = -0.0003$~meV, $D_y =-0.025$~meV, $D_z=-0.006$~meV, and a Lorentzian broadening of $\Delta E = 0.73$~meV for a combined reduced $\chi^2$ of 17.1. We note that there is no attempt to quantify any errors on the best-fit parameters, since systematically mapping out the six dimensional $\chi^2$ landscape related to the fit is beyond the scope of the article.

The fit captures the dominant features of the spin waves quite well, as is especially evident in Fig.~\ref{fig:SW}(c) of the actual fit. The fit highlights that the system is highly two-dimensional, since $J_\perp$ is not significantly different from zero. We note that $J_2$ is required to satisfactorily model the spin waves. A key feature of the fit is the Lorentzian broadening in energy, $\Delta E = 0.730$~meV. This is 20 times larger than the instrumental resolution of the elastic line, $\delta E = 0.082$~meV, and will be discussed in the detail in section \ref{sec:disc-SW}. 
When the ground state is optimized, SpinW achieves a 120$^\circ$ spin structure as expected with a spin canting angle of 1.06$^\circ$ out of the kagom\'e plane. This matches well with the canting angle estimated from our magnetization measurements of 0.7$^\circ$. Furthermore, the sum of all exchange interactions from SpinW matches well with the mean-field  $J_{\mathrm{MF}} \approx 1.0$~meV estimate from the susceptibility data.

\subsection{Diffuse excitations above \texorpdfstring{$T_N$}{TN}}
For $T > T_N$ the diffuse excitations in the range 0-3~meV persist, and above $\sim 35$~K ($> 9~ T_N$) the signal resembles a simple, paramagnetic background. 
We will attempt to model the excitations above $T_N$ by describing the excitations along the energy and Q axes, separately.

\subsubsection{Energy dependence}
\label{sec:E}
\begin{figure}[b]
    \includegraphics{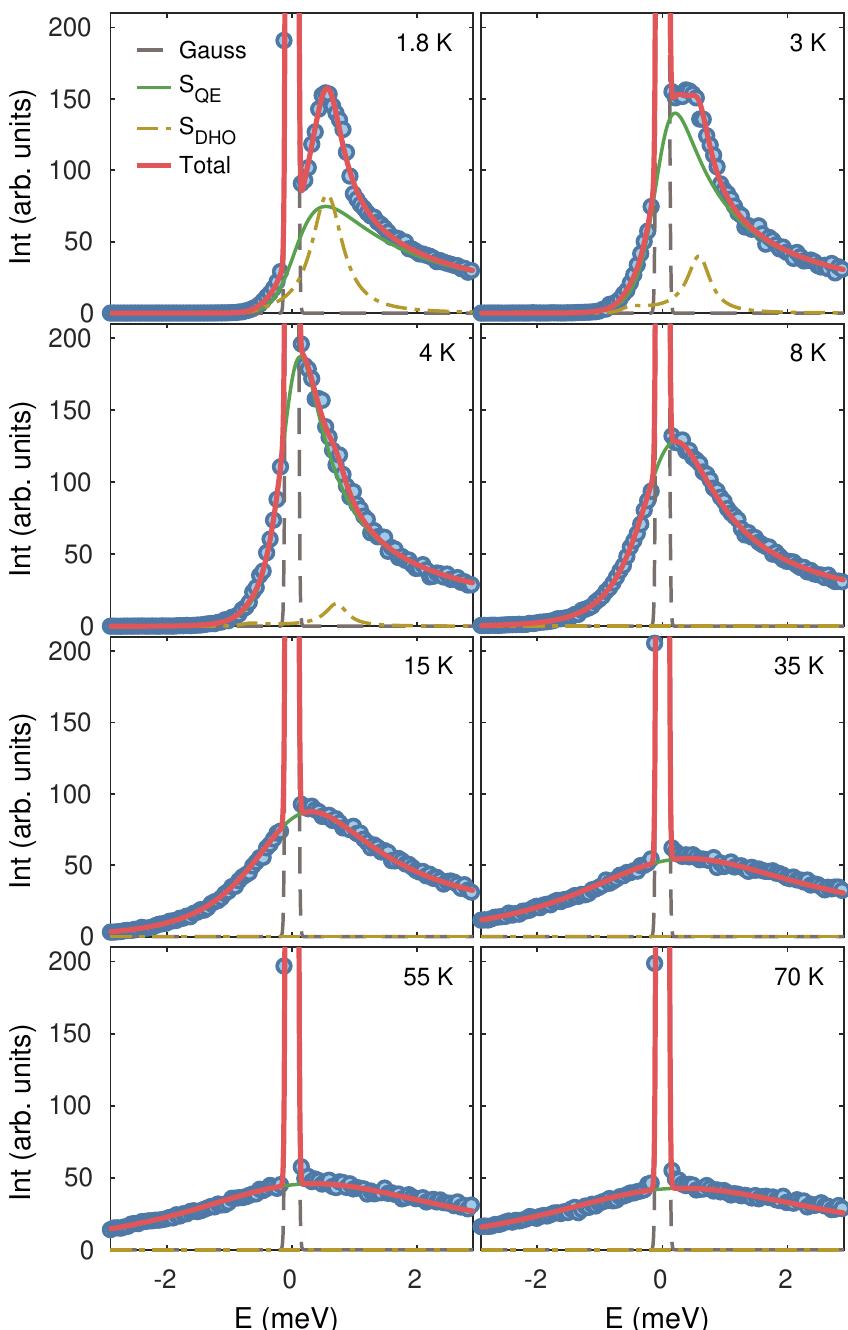}
    \caption{Constant-Q cuts at 1.1 $\pm 0.1$ \iang fitted with the damped harmonic oscillator model of eq.~\ref{eq:DHO}.} 
    \label{fig:DHO}
\end{figure}

The temperature evolution of constant-Q cuts around the magnetic Bragg reflection at 1.1 \iang is shown in Fig.~\ref{fig:DHO}. 
The data can be modeled assuming a quasielastic behavior with broadening in energy due to finite life-times and a damped harmonic oscillator (DHO) to account for the inelastic response at low temperatures \cite{Hayashida2020,Faak1997}. Thus, we fit the data with the following model:
\begin{align}
    S(\omega,T)=S_E(\omega,T)+S_{\rm QE}(\omega,T)+S_{\mathrm{DHO}}(\omega,T),
    \label{eq:DHO}
\end{align}
where $S_E(\omega,T)$ is a Gaussian to account for the elastic line. $S_{\rm QE}(\omega,T)$ is the quasielastic response as modeled by an exponential spin relaxation in the form of a Lorentzian
\begin{align}
    S_{\rm QE}(\omega,T)=\frac{1}{1-e^{-\hbar \omega / k_{\mathrm{B}} T}} \frac{\chi_{0} \omega \Gamma}{\omega^{2}+\Gamma^{2}},
\end{align}
where the first term represents the detailed balance factor describing the thermal population of excited states, $\Gamma$ is the energy line width, and $\chi_0$ is the static susceptibility. 
The inelastic response $S_{\mathrm{DHO}}(\omega,T)$ is fitted by a damped harmonic oscillator (DHO) including detailed balance factor and corresponding to the double Lorentzian
\begin{align}
    S_{\mathrm{DHO}}(\omega,T)=  \frac{1}{1-e^{-\hbar \omega / k_{\mathrm{B}} T}} \frac{A_{\mathrm{DHO}} \omega \Gamma}{\left(\omega^{2}-\omega_{\mathrm{DHO}}^{2}\right)^2+
    \left(\omega \Gamma \right)^{2}},
\end{align}
where $A_{\mathrm{DHO}}$ is the oscillator strength and $\omega_{\mathrm{DHO}}$ is the eigenfrequency. $S_{\mathrm{DHO}}(\omega,T)$ is only included in the fit for 4~K and below, since the scattering above could easily be reproduced without the inelastic response.

The fit to data is shown in Fig.~\ref{fig:DHO}, and shows that the inelastic response at low temperatures is pronounced, but decreases steadily as the temperature increases. Simultaneously, the energy line width $\Gamma$, which is common to both $S_{QE}(\omega,T)$ and $S_{\mathrm{DHO}}(\omega,T)$, increases as the temperature increases as expected. It should be noted that the line width at 1.8~K of $\Gamma = 0.6 \pm 0.2$ meV is consistent with the energy broadening obtained by the SpinW model of 0.73 meV.

\begin{figure}[tb]
    \includegraphics{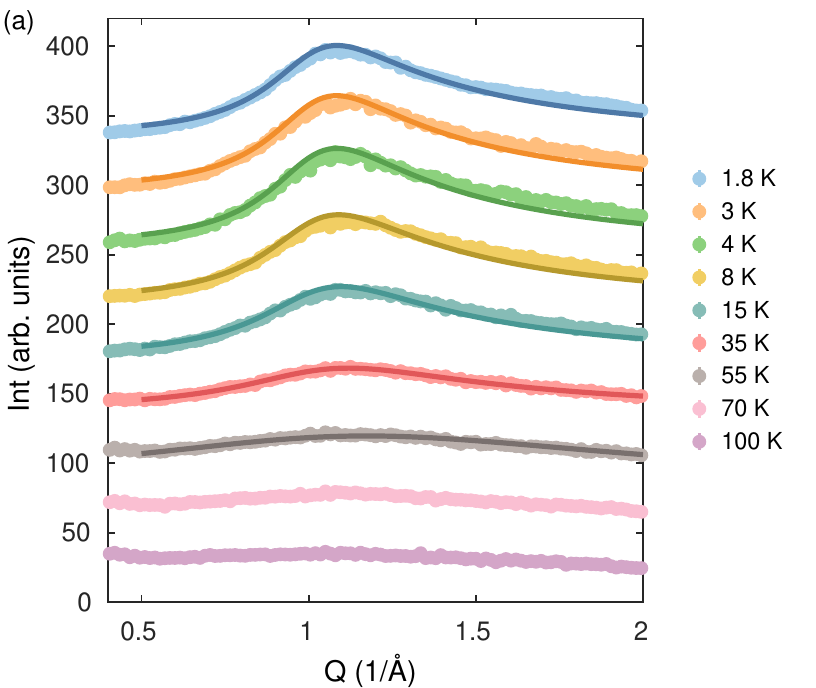}
    ~
    \includegraphics[trim= 0 0 -1.3cm 0,clip]{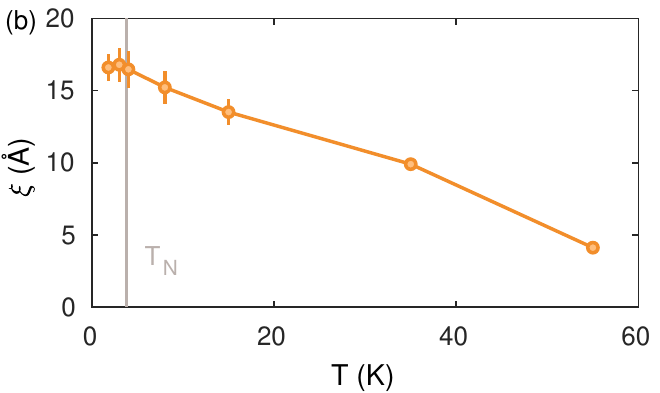} 
    \caption{(a) Q-dependence of data integrated in the range $0.11-2.5$~meV for various temperatures. Full lines are fit to eq.~\ref{eq:cluster}. Each temperature has been offset for clarity. (b) Resulting correlation length.}
    \label{fig:cluster}
\end{figure}

\subsubsection{\textit{Q}-dependence}
\label{sec:Q}
The temperature evolution of the Q-dependence of the excitations may reveal information about the short-ranged nature of the correlations. Fig.~\ref{fig:allT} shows the Q-dependence of the excitations for various temperatures obtained by integrating the spectroscopy data for excitations above the elastic line. The data shows a very broad and asymmetrical peak at $Q\approx 1.1$\iang. The diffuse appearance and the broadness in $Q$ of the excitations indicate short-ranged magnetic correlations. These excitations are strongest for lowest temperatures, but persist until around $35 K \approx |\theta_{\rm CW}|/2$, much above $T_N$. Due to the dominating in-plane couplings between spins, we hypothesize that this behavior arises from short-ranged correlations that are purely two-dimensional and occur between spins within the kagom\'e planes. In order to model this, we employ a heuristic model for obtaining powder-averages of 2D excitations with characteristic length scale $\xi$.\\

\begin{table*}[ht]
\centering
\begin{tabular}{l l c c c c c c c}
 & Method & $J_1$ & $J_2$ & $J_\perp$ &  $D_y$ & $D_z$ & $\Delta E$ & Canting\\ \hline \hline
\citet{Lee1997} & Susceptibility & 1.21  & --  & -- & -- &--  &-- &--  \\ \hline
\citet{Okuta2011} &  Susceptibility & 1.06 & -- & -- &  -- &   --  &--  &	0.86$^\circ$ 	 \\ \hline
\citet{Okubo2017} & Antiferromagnetic resonances & -- &  -- & -- &  -0.0470  & -0.0120  &-- &  1.44$^\circ$   \\ \hline
This article  & Neutron spectroscopy & 0.881 & 0.015 & -0.0003 & -0.025 &  -0.006 & 0.73 & 1.06$^\circ$  \\ \hline 
\end{tabular}
\caption{Exchange couplings, energy broadening and spin canting for Cr-jarosite estimated in the literature and from the SpinW simulation of our data based on the Hamiltonian in eq.~\ref{eq:jaroham}. All values are given in units of meV, unless otherwise stated.}
\label{tab:crjaropar}
\end{table*}

In this model, we assume that the 2D excitations will occur around the 2D ordering vector $(hk)=(10)$ located at $Q_0=1.005$\iang, but with constant intensity independent of the $l$ component of {\bf Q}. This 2D scattering rod must then be powder-averaged in three-dimensional $(hkl)$ space. The geometrical derivation of this can be found in appendix \ref{app:geometry}, and yields an effective expression for the powder-averaged Q-dependence of the intensity as $I(Q) \propto 1/\left(Q\sqrt{Q^2-Q_0^2}\right)$ for $Q>Q_0$ and zero for $Q < Q_0$. The divergence at $Q=Q_0$ is essentially of a square root type and thus integrable.
Our result is identical to the treatment of \citet{Yang1996}, which is an improvement of the classical Warren function \cite{Warren1941}, valid for powder-averaged diffraction of a 2D system.
To account for the short-ranged correlations of our magnetic system, we perform a convolution of the expression with a Lorentzian with a full width at half maximum given by $\gamma = 1/\xi$, where $\xi$ is the correlation length. Thus, the resulting convoluted line shape is
\begin{align}
S(Q,T) &= C_L(Q,T) = |f(Q)|^2 \times \label{eq:cluster} \\
&\int_{0}^{\infty} \frac{1}{Q^\prime\sqrt{{Q^\prime}^2-Q_0^2}} ~\frac{\gamma^2}{(\gamma/2)^2 + (Q^\prime-Q)^2}~ dQ^\prime,\nonumber
\end{align}
where $f(Q)$ is the magnetic form factor for \ce{Cr^{3+}}. The equation is fitted to the data including an overall scaling parameter and a variable background. In our data analysis, $\gamma$ is sufficiently broad that convolution with an experimental resolution function is unnecessary.

The fitting of this model to the data can be seen in Fig.~\ref{fig:cluster}, which shows the Q-dependence of the data integrated from $0.11 - 2.5$~meV, {\em i.e.}\ the excitations above the elastic line. Temperatures above 55~K $\approx |\theta_{CW}|$ have not been fitted due to lack of distinct features. For the lower temperatures, our simple model works remarkably well, as it reproduces both the peak position and asymmetric line shape; both in the classical spin liquid regime and below $T_N$. 
The temperature dependence of the effective 2D correlation length $\xi$ is shown in Fig.~\ref{fig:cluster}(b). At the highest fitted temperature, 55~K, we find $\xi \approx 4$~\AA~ which is close to the distance between nearest-neighboring Cr atoms in the kagom\'e plane of 3.65~\AA. The correlation length increases for decreasing temperature, and reaches $\xi \approx 17$~\AA\ at 1.8~K, which corresponds to 4.6 times the separation distance of the Cr atoms.


\section{Discussion}

\subsection{Effects of the sample quality}
\label{sec:disc-quality}
Our sample has 2.8\% magnetic \ce{Cr^3+} vacancies. This is significantly closer to stoichiometry than the only other sample measured with neutron spectroscopy, which had 26\% vacancies \cite{Lee1997}. Furthermore, our sample has 14--20\% \ce{K+} vacancies that may either be replaced by \ce{D2O} or \ce{D3O+}. \citet{Lee1997} did not quantify this value, but it is unlikely that their K vacancies are significantly smaller than ours. Thus, this study is the best current measurement of excitations in Cr-jarosite.

For the iron analogue jarosites, it is well-established that \ce{(H3O)Fe3(SO4)2(OH)6} is a spin glass \cite{Wills2001xyz,Grohol2007,Spratt2014}, and furthermore, the magnetic behavior changes gradually between the $120^\circ$ AFM order in \ce{KFe3(SO4)2(OH)6} and the spin glass behaviour of \ce{(H3O)Fe3(SO4)2(OH)6} with increasing \ce{H3O+} content \cite{Grohol2007}. For the Cr analogues, the precise magnetic behavior of \ce{(H3O)Cr3(SO4)2(OH)6} remains largely unexplored, although it also shows hints of spin glass behavior \cite{WillsPhD,Mendels2011,Wills2001}. Assuming that hydronium defects act in a similar manner in both Cr- and Fe analogues, the level of K vacancies in our sample will certainly have some effect on the magnetic ordering. Assuming that the vacancies are approximately randomly located throughout the lattice, this could lead to a distribution of exchange interactions. The consequence of this will likely be increased levels of diffuse scattering, which is hard to distinguish from what is caused by the frustration in the kagom\'e lattice.

\subsection{Spin waves and coupling constants}
\label{sec:disc-SW}
The resulting exchange parameters as obtained from powder-averaged linear spin wave theory are summarized in table \ref{tab:crjaropar}, where they are compared with values from the literature. Our exchange parameters improve on the previously published estimates of the exchange couplings, as our model is the first to include all the interactions simultaneously. The value of the nearest neighbor coupling parameter, $J_1$, is around 17\% smaller than previously estimated by magnetization measurements \cite{Okuta2011}. However, our inelastic neutron spectroscopy is expected to be a much more precise estimator. The DM coupling we obtain via SpinW is almost 50\% lower than determined by electron spin resonance (ESR) \cite{Okubo2017}. However, we expect ESR to provide order of magnitude estimate, and in this regard it matches our DM interactions well. The spin wave spectrum calculated with the same parameters but using the DM couplings suggested by Ref.~\cite{Okubo2017} decrease the match with data significantly. Furthermore, our results indicate that $J_\perp$ does not differ significantly from zero within the uncertainty of the fit. 
Furthermore, the spin canting angle of 1.06$^\circ$ obtained by the SpinW model agrees with both our estimate from susceptibility of 0.7$^\circ$, as well as the canting angles measured in the literature in the range $0.86-1.44^\circ$ \cite{Okuta2011,Okubo2017}.

The spin waves appear diffuse in both $Q$ and $\hbar \omega$. A number of possible, non-exclusive reasons for this exist: it could indicate a broadening due to finite life times of the excitations, caused by the frustration in the lattice or by the relatively low spin value ($S=3/2$); or it could indicate a distribution of exchange interactions due to randomly located vacancies at the A-site. Our spinW model takes this into account in an effective way by including the Lorentzian broadening, $\Delta E$. 

The intense spin wave feature at $Q\approx 1.1$\iang has an incomplete gap, but the flat mode has a full gap of $\sim 0.2$~meV. Both gaps close at $T_N \approx 3.8$~K. Our measurement thus provides evidence of the flat kagom\'e zero mode lifted to finite energies by the DM interaction, so far only observed in iron jarosite \cite{Matan2006}, in the compound \ce{NaBa2Mn3F11}  \cite{Hayashida2020}, and indirectly via magnetic resonance in \ce{Li9Fe3(P2O7)3(PO4)2} \cite{Kermarrec2021}.

\subsection{Classical spin liquid phase}
\begin{figure}[htbp]
    \includegraphics[trim = {-0.3cm 3.6cm 0 0},clip]{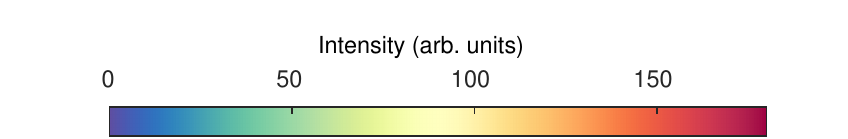}
    \includegraphics{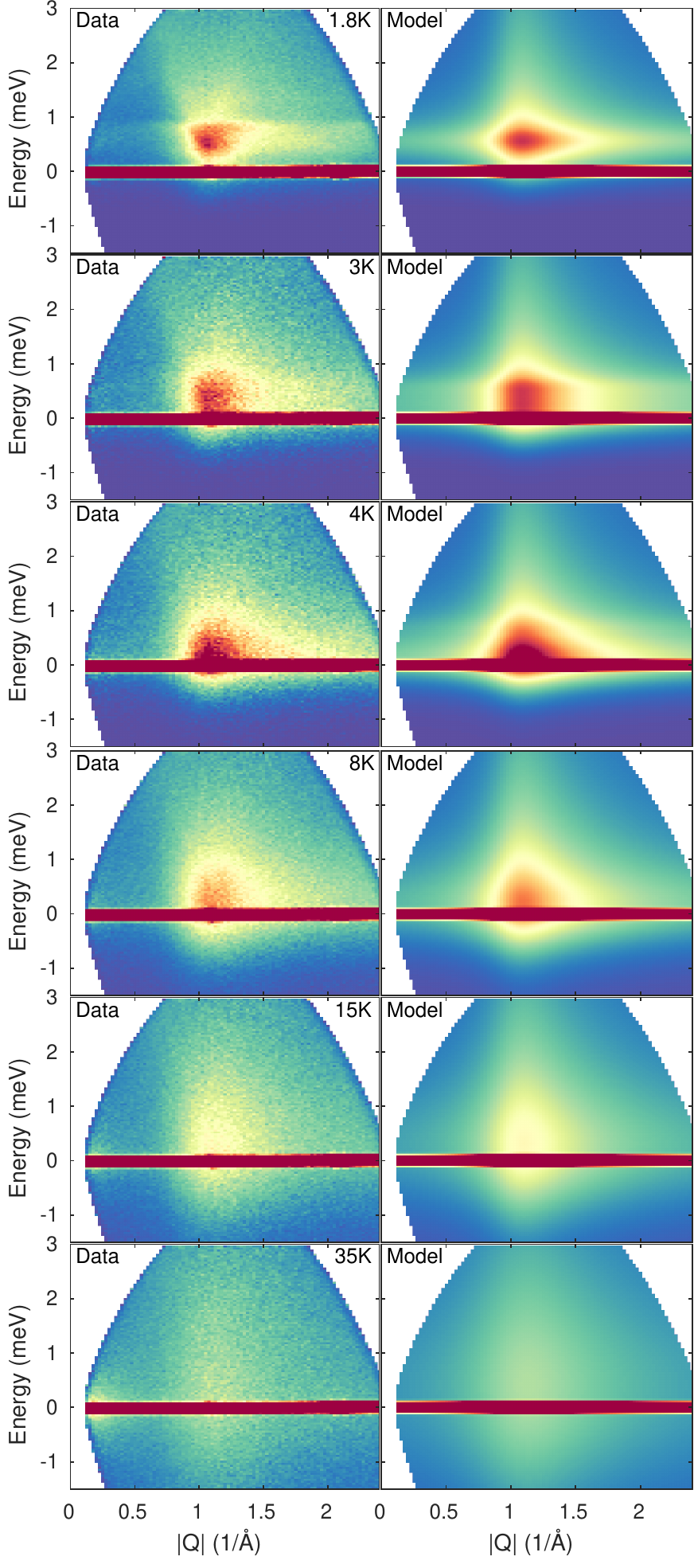}
    \caption{Left column: raw LET data for selected temperatures. Right column: Normalized $S_{\mathrm Fac}(Q,\omega,T)$ (eq.~\ref{eq:Qxw} on a common color scale.} 
    \label{fig:fac1}
\end{figure}

Having obtained working models of the energy- and $Q$-dependence of the excitations, it is natural to investigate whether we can combine these for a complete description. One possibility, inspired by ref.~\cite{Bai2019}, is a factorization of the dynamical response, which implies that the energy- and $Q$-response is fully decoupled in the system. In such a system, small spin clusters would fluctuate independently of each other. The factorization combines the two theoretical descriptions obtained in sections \ref{sec:E} and \ref{sec:Q} to read:
\begin{align}
    S_{\mathrm Fac}(Q,\omega,T) = S(Q,T) \times S(\omega,T). \label{eq:Qxw}
\end{align}
The results of this analysis is displayed in Fig.~\ref{fig:fac1}, which compares the raw LET data to the model resulting from sections \ref{sec:E} and \ref{sec:Q}. Here, $S_{\mathrm Fac}(Q,\omega,T)$ has been normalized, such that the sum of intensity over $Q$ and $\omega$ above the elastic line matches that of the experimental data for each temperature. \\ 

As expected, the factorization model fails below $T_N$ (Figs. \ref{fig:fac1}, 1.8~K and 3~K): it does not capture the spin waves, which is intrinsically a phenomenon with coupled energy- and $Q$-response. In contrast, at the highest shown temperature of 35~K, the factorization model works quite well (apart from the structural feature at very low $Q$). This is to be expected,  since the system is now close to a simple, paramagnetic response with decoupled spin clusters fluctuating independently. At intermediate temperatures of 4~K and 8~K, the factorization also does not capture essential features of the data, indicating that the system remains dominated by strongly coupled excitations that are correlated in energy- and $Q$-response. This is in line with the system being in its classical spin liquid regime at these temperatures.
This is in agreement with observations in other frustrated compounds, including the pyrochlore \ce{MgCr2O4} \cite{Bai2019}, which also finds that factorization fails in the classical spin liquid regime.

This overall behavior is similar to recent observations in a similar compound, the classical spin liquid $h$-\ce{YMnO3}, which is a triangular-lattice Heisenberg magnet with $S=2$ \cite{Janas2021}. These single-crystal neutron studies showed that small 2D clusters fluctuate in the classical spin liquid regime, and coexist with the spin waves below the 3D ordering transition in a non-interacting manner. As the temperature is decreased significantly, the spin waves increasingly dominate the picture. 

Thus, for Cr-jarosite, the diffuse appearance of the spin waves could also stem from a coexistence and superposition of the 2D dynamics appearing in the classical spin liquid regime above $T_N$. This could also explain the lack of a complete spin wave gap at above the (101) magnetic peak at 1.8~K, if 2D scattering is superimposed on the spin waves. This is supported by  the fact that the spin waves in the ordered state, at 1.8~K~$\sim T_{\rm N}/2$ appear strongly damped, as judged from the large Lorentzian broadening obtained by the modeling. Of course, these effects are less clear in powdered samples, and we highlight the need for future single-crystal studies to investigate these phenomena further.

\FloatBarrier

\section{Conclusion}
We have measured the excitations in well-characterized powder samples of Cr-jarosite. Neutron spectroscopy at low temperatures revealed diffuse spin waves with kagom\'e zero mode excitations, which were modeled with linear spin wave theory to obtain estimates for the exchange couplings.
Persistent, diffuse excitations above $T_N$ in the classical spin liquid regime were also observed. The energy-response of these were modeled assuming an exponential spin relaxation as well as a damped harmonic oscillator model representing the inelastic response below $T_N$. The $Q$-response were modeled based on 2D cluster excitations. The energy- and $Q$-response was found to be strongly coupled in both the ordered and classical spin liquid regime, until it decoupled in the fully paramagnetic phase closer to $\theta_{\rm CW}$. This is indicative of classical spin liquid behavior.

\section*{Acknowledgments}
We thank the ISIS neutron facility, Rutherford Appleton Laboratory (United Kingdom) for awarding us beam time (experiment RB1990088). The experimental work was supported by the Danish Foundation for Independent Research through the DANSCATT (Grant No. 7055-00010B) and the Research Council for Natural Sciences (U.G.N., A.B.A.A. and S.L.L.; Grant No. DFF-FNU-7014-00198). 
We thank Machteld E. Kamminga for insight and expertise into Rietveld refinements, H\o gni Weihe for helpful discussions on ESR, and Christian B. J\o rgensen for assistance with the NMR measurements. 
\bibliography{Mybib}

\appendix

 \section{Synthesis details}
 \label{app:synth}
 \ce{KCr3(OD)6(SO4)2} was synthesized using the redox-hydrothermal method. 4.65 g of \ce{K2SO4} (99.0 \%, Sigma Aldrich) was dissolved in 42 mL \ce{D2O} (98 \%, Cambridge isotopes laboratories), 2 mL of concentrated sulfuric acid (18 M) was added and the solution  transferred to a 80 mL teflon lined stainless steel hydrothermal reaction vessel, which contained  0.67 g \ce{Cr}   (99.5 \%, Sigma Aldrich) pellets were placed in a Binder oven.  The reaction vessel was heated at 210 $^\circ$C for 5 days for samples 1 to 4 whereas a reaction time of 7 days were used for samples 5 and 6. The product was isolated by filtration, washed  with \ce{D2O}, and dried in a oven at 60 $^\circ$C for 4-24 hours.
 The details of the synthesis is summarized in table \ref{tab:synthesis}
 
\begin{table}[htb]
\begin{tabular}{ccccccc}
\hline
Sample  &  \ce{K2SO4} &   Cr  &  \ce{D2O} &   \ce{H2SO4} &   Yield &  Yield  \\
no. & (g) & (g) & (mL) & (mL) & (g) & (\%) \\ \hline
	1  &  4.6596 &   0.6835 &   42 & 2 & 1.5052 & 77  \\
	2  &  4.6578 &   0.6960 &   42 & 2 & 1.4238 & 72  \\
	3  &  4.6710 &   0.6775 &   43 & 2 & 1.4403 & 75  \\
	4  &  4.6657 &   0.6760 &   42 & 2 & 1.6017 & 83  \\
	5  &  4.6670 &   0.6715 &   42 & 2 & 1.72   & 90  \\
	6  &  4.6818 &   0.6740 &   42 & 2 & 1.6    & 84  \\ \hline
\end{tabular}
\caption{Amount of reactants used in each synthesis, yield of each synthesis, and percentage of the theoretical yield. 
}
\label{tab:synthesis}
\end{table}

 \section{PXRD}
 \label{app:PXRD}
\begin{figure}[h]
    \includegraphics[width=\columnwidth]{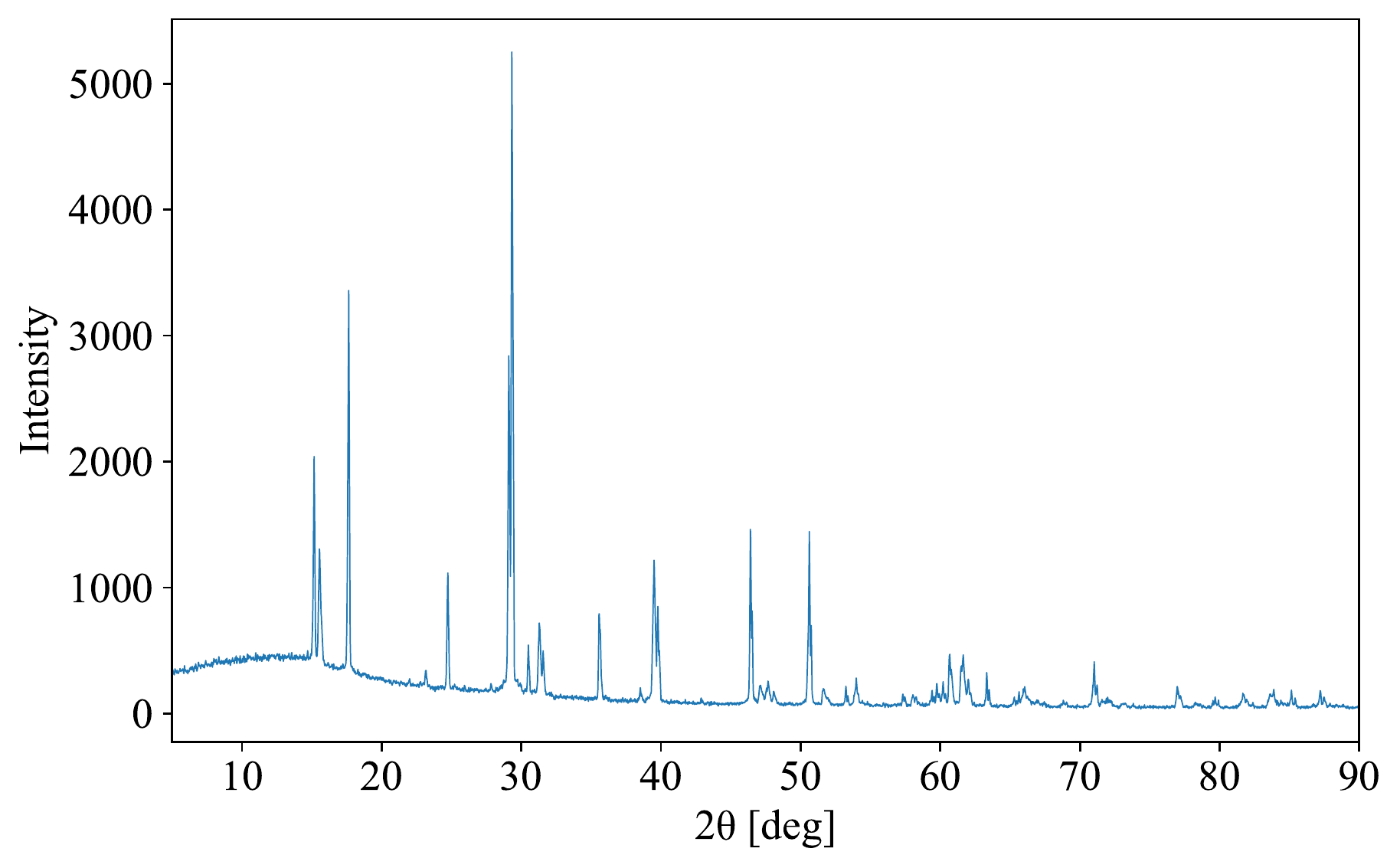}
    \caption{PXRD spectrum of combined sample, the broad background is from the instrument.} 
    \label{fig:PXRD}
\end{figure}

The PXRD diffractograms of all samples were recorded on a Rigaku Miniflex using a copper target (K$_\alpha$ = 1.5405~\AA) and collected in range $5 - 90^\circ$ (2$\theta$) with a step size of 0.02$^\circ$ and a scan rate of 10$^\circ$ pr min. Fig.~\ref{fig:PXRD} shows the PXRD diffractogram of the Cr-jarosite sample. This matches earlier reported diffractograms \cite{Janas2020,Grube_2018}. 

\section{SEM}
 \label{app:SEM}
 \begin{figure}[htb]
    \includegraphics[width=0.4\textwidth]{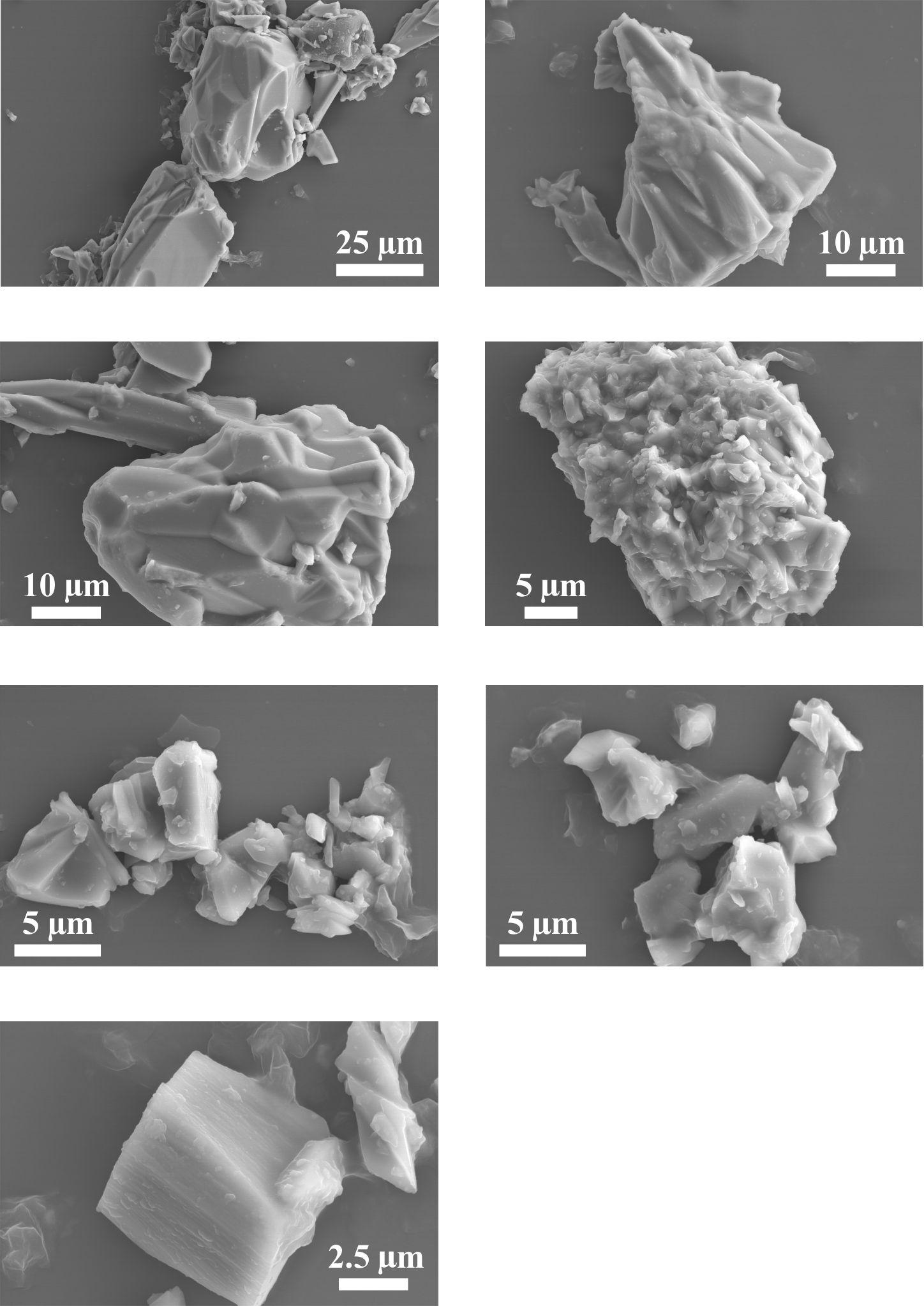}
    \caption{SEM images of a Cr-jarosite sample at magnifications 25 $\mu$m to 2.5 $\mu$m illustrating the particle morphology.} 
    \label{fig:SEM}
\end{figure}
 Selected samples including the combined Cr-jarosite were characterized by Scanning Electron Microscopy (SEM) with Energy Dispersive Spectroscopy (EDS) for quantitative analyses of the heavier elements (K, S, and Cr) and sample morphology. EDS and SEM was performed on well-dispersed samples as this allowed us to probe the composition and surface of the individual crystallites, as the primary focus was on sample homogeneity. However, it should be noted that the precision of EDS is improved when measured on the smooth surface of pressed pellets, so the EDS measurement has increased uncertainty.
 
 The measurement was performed on a Hitachi S-4800 scanning electron microscope equipped with an EDS. Approximately 1 mg sample was dispersed into 1 mL of ethanol by ultrasonification for 15 min. A drop of the suspension was placed on a silicon wafer and then dried at 50$^\circ$ C for 10 min. Subsequently, the surface of the sample was coated with an approximately 15 nm layer of gold. EDS was measured in five random areas of the sample grid using a 15 kV accelerated electron beam and 15 mm working distance. Only sample batches 2 and 3  were measured. 

Figure \ref{fig:SEM} shows the sample at different magnification levels illustrating the particle morphology. The measurements reveal both irregularly shaped and aggregated particles of size 10-50~$\mu$m, in agreement with the stacking disorder reported in the earlier study \cite{Janas2020}. No additional phases were observed by SEM. The EDS measurements of individual Cr-jarosite crystallites yield a K:Cr:S ratio of 0.99(1):3.3(3):2.00(7) normalized to an S occupancy of 2. This is rough agreement with the stoichimetric 1:3:2 ratio taking the increased experimental uncertainties into consideration. The \ce{Cr} content is slightly above the expected, although with large uncertainties. It may potentially be ascribed to the presence of smaller pieces of unreacted chromium metal in the samples, or reflect large uncertainties of the EDS measurements.We note that several larger chromium metal pieces were present in the isolated product and removed manually before the samples were dried. Therefore, we carefully inspected the PXRD, NMR, and neutron data to look for indications of remaining unreacted Cr. PXRD and neutron data showed no reflections from metallic Cr, and no indication of the antiferromagnetic phase transition at 311 K for Cr is observed in the susceptibility data (see Fig.~\ref{fig:MPMS}a). Thus, if Cr impurities are present these are negligible. 

 Based on the experimental uncertainties, the results from \ce{^{2}H} MAS NMR provides a more precise quantitative assessment of the Cr and K contents than EDS and PXRD, and we will use the \ce{^{2}H} MAS NMR results as the most precise estimate going forward.

 \section{Simulation of the \texorpdfstring{\ce{^2}H}{2H} MAS NMR spectrum}
 \label{app:NMR}
\begin{figure}[htb]
    \includegraphics[width=0.4\textwidth]{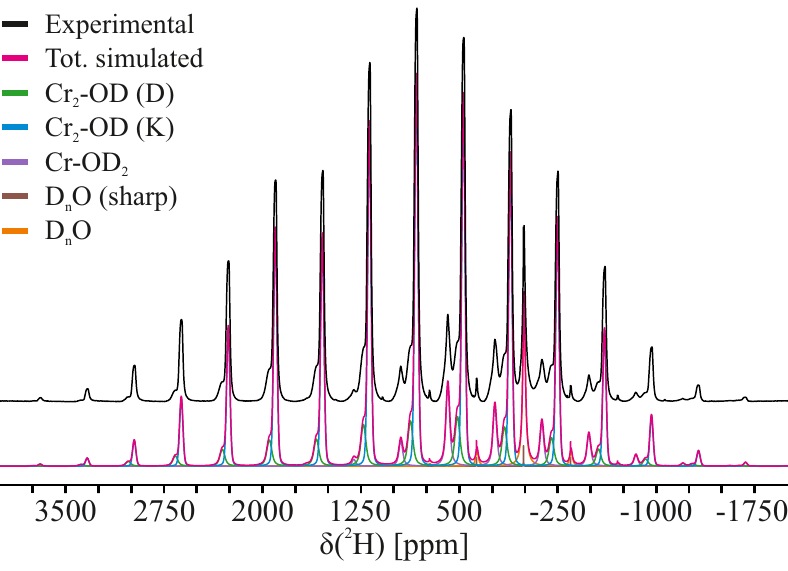}
    \caption{Experimental and simulated \ce{^{2}H} MAS NMR spectrum (Hahn Echo) of 
    combined sample recorded with 33 kHz spinning speed using the parameters in Table \ref{tab:NMR-properties} .} 
    \label{fig:NMR}
\end{figure}

Figure \ref{fig:NMR} illustrates the experimental and simulated \ce{^{2}H} MAS NMR spectrum of the combined Cr-jarosite sample using the the parameters obtained from fitting of the experimental data in Table \ref{tab:NMR-properties}. 

\begin{table*}[htb]
\begin{tabular*}{\textwidth}{l @{\extracolsep{\fill}} lllllll}
\hline
Site & $\delta_{iso}$ {[}ppm{]} & $\Delta$ {[}ppm{]} & $\eta_{\sigma}$ & $C_Q$ {[}MHz{]} & $\eta_Q$ & $\beta$ {[}deg{]} & Intensity {[}\%{]} \\ \hline
\ce{Cr2}-\ce{OD} (D) & 872(10) & -1020(100) & 1.00(10) & 0.216(40) & 0.10(10) & 45(10) & 15(1)           \\
\ce{Cr2}-\ce{OD} (K) & 828(5)  & -935(50)   & 1.00(10) & 0.214(10) & 0.09(10) & 55(5)  & 69(1)           \\
\ce{Cr2}-\ce{OD}     & 229(5)  & 830(50)    & 0.00(10) & 0.101(5)  & 0.64(10) & 74(10) & 10(1)           \\
D$_n$O (sharp)       & 11(1)   & 0(10)      & 0.00(10) & 0.100(5)  & 0.00(10) & 0(5)   & 1\textgreater{} \\
D$_n$O               & 7(2)    & 0(10)      & 0.00(10) & 0.050(5)  & 0.00(10) & 0(5)   & 6(1)            \\ \hline
x:    & 0.085(5)     & Cr [\%]:  & 2.8(3) & n:         & 0.41(10) & K [\%]:        & $\sim$ 14-20        \\ \hline
\end{tabular*}
\caption{\ce{^{2}H} MAS NMR parameters for the 
combined sample recorded at 33 kHz spinning speed. The CSA parameters ($\Delta$ and $\eta_{\sigma}$), quadrupole coupling ($C_Q$ and $\eta_Q$), signal intensities (Int.), Cr-site and K-site vacancies were found by fitting of the recorded \ce{^{2}H} MAS NMR spectrum.\cite{ssnake} The CSA parameters are given in the Haeberlen convention,\cite{Haeberlen} where $\beta$ is an Euler angle between principal axis of the CSA and quadrupole tensor.\cite{skibsted1992} The vacancies are given per mol. The errors were estimated from the upper and lower boundaries of the fitted parameters.}
\label{tab:NMR-properties}
\end{table*}

\section{Derivation of Q-dependent model}
\label{app:geometry}
\begin{figure}[htb]
{{\includegraphics[width=0.9\columnwidth]{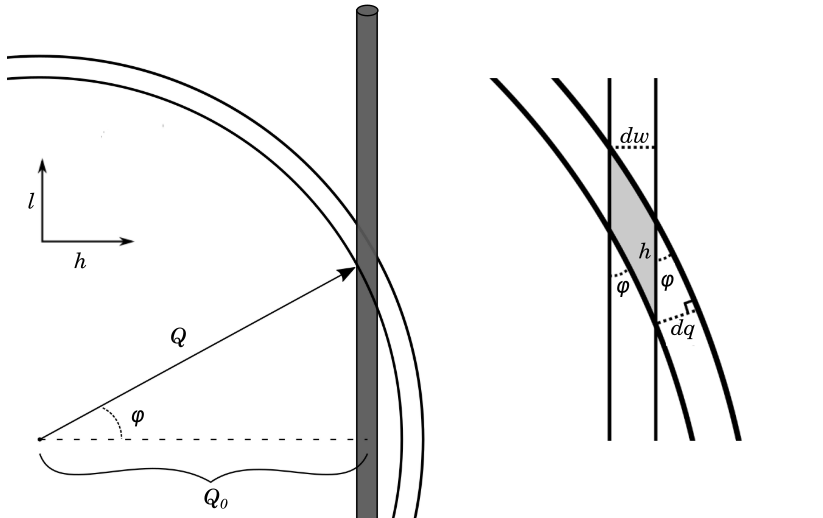}}}
\caption{Reciprocal lattice representation of scattering by 2D reflections. 
}
    \label{fig:poweq-app}
\end{figure}

We hypothesize that the excitations above $T_N$ originate from two-dimensional dynamics of correlated spin clusters within the kagom\'e planes. Thus, we assume that the 2D excitations will be centered at the 2D ordering vector $(hk)=(10)$, located at $Q_0=1.005$\iang. Due to the lack of 3D correlations, this scattering will occur as scattering rods around {\bf Q}$ = (1 0 l)$. To obtain the observed pattern, the rods must then be powder-averaged in three-dimensional $(hkl)$ space.
This is sketched geometrically in Fig.~\ref{fig:poweq-app}. Here the shaded area represents the overlap between the sphere of $(hkl)$ points of constant length, $Q$, and the cylindrical rod at $(10l)$, which has its smallest $Q$-value at $(100)$, with the length $Q = Q_0$. The shape of the scattering intensity is thus related to the overlap volume, which must be calculated.

The $(hk)$ rod has width $dw$, and the $Q$-sphere has width $dq \ll Q$ as illustrated in figure \ref{fig:poweq-app}. Assuming that the intersection is not curved appreciably over the area, then the overlap area is $A=h~dw$, where $h$ is the effective height. $h$ is also the hypotenuse in the triangle as indicated, and can be written as $h = dq / \sin\phi$, which thus makes the area $A=dw~dq/\sin\phi$.
By realizing that the $\phi$ is equivalent in the triangles sketched in the left and right panels in figure \ref{fig:poweq-app}, then it can be related to the scattering vectors via $\cos\phi = Q_0/Q$. 

The scattering intensity can be found as the ratio between the overlap volume, $V_o = A~dw$ and the volume of the spherical shell, $V_s = 4\pi Q^2dQ$:
\begin{align} \label{eq:yang}
\mathcal{V}(Q) = \frac{V_o}{V_s} = \frac{dq~dw^2 / \sin\phi}{4 \pi dq Q^2} \propto \frac{1}{Q\sqrt{Q^2-Q_0^2}}, 
\end{align}
where the constants can be accounted for in an overall scaling factor. Now, this expression is only valid for $Q>Q_0$, and takes a value of zero for $Q<Q_0$. The expression diverges at $Q=Q_0$ in a square root manner, making the integral $\int \mathcal{V}(Q) dQ$ take a finite value. This expression was also found in Ref.~\cite{Yang1996}, which also finds a modification to eq.~(\ref{eq:yang}) close to the divergence that is unnecessary for our purpose. These results are, in turn, a correction of the classical results for 2D line shapes by Warren \cite{Warren1941}.

A way to account for the short-range correlations in the magnetic system is to assume a finite-size broadening of the value of $Q_0$. We obtain this by a convolution of $\mathcal{V}(Q)$ with a Lorentzian to model smearing by short-ranged correlations with correlation length $\xi=1/\gamma$, where $\gamma$ is the full-width-half-max of the Lorentzian. Thus, the resulting convoluted line shapes is
\begin{align}
C_L(Q) =& |f(Q)|^2 \times \\ &\int_{0}^{\infty} \frac{1}{Q^\prime\sqrt{{Q^\prime}^2-Q_0^2}} ~\frac{\gamma^2}{(\gamma/2)^2 + (Q^\prime-Q)^2}~ dQ^\prime.\nonumber
\end{align}
Here the magnetic form factor $f(Q)$ for \ce{Cr^3+} has been added using tabulated values from ref. \cite{formfactor}. The intensity has zero value for small values of $Q$, but starts to increase when the Lorentzian just touches $Q_0$. As $Q$ approaches $Q_0$ the value increases rapidly. At values clearly larger than $Q_0$, the value slowly decays. Thus, the asymmetric peak shape is caused by the asymmetry of the overlap function $\mathcal{V}(Q)$. As an additional result of the convolution, the peak position is displaced to a value of $Q$ slightly larger than $Q_0$.

\end{document}